\documentclass[aps,pre,twocolumn]{revtex4}
\usepackage{amssymb}

\usepackage{eurosym}
\usepackage{epsfig}

\begin{document}

\title{Isotropization of two-dimensional hydrodynamic turbulence in the direct cascade}

\author{E.A. Kuznetsov$^{a,b,c}$\/\thanks{%
kuznetso@itp.ac.ru} and E.V. Sereshchenko$^{a,d,e}$}
\affiliation{{\small \textit{$^{a}$ Novosibirsk State University, 630090 Novosibirsk,
Russia} }\\
{\small \textit{$^{b}$ Lebedev Physical Institute, RAS, 119991 Moscow, Russia%
}}\\
{\small \textit{$^{c}$ Landau Institute for Theoretical Physics, RAS, 119334
Moscow, Russia}}\\
{\small \textit{\ $^{d}$ Khristianovich Institute of Theoretical and Applied
Mechanics, SB RAS, 630090 Novosibirsk, Russia}}\\
{\small \textit{$^{e}$ Far-Eastern Federal University, 690091 Vladivostok,
Russia }}}

\begin{abstract}
We present results of numerical simulation  of the direct cascade in two-dimensional hydrodynamic turbulence (with spatial resolution up to $16384 \times 16384$). If at the earlier stage (at the time of order of the inverse  pumping growth rate $\tau\sim\Gamma_{max}^{-1}$), the turbulence develops according to the same scenario as in the case of a freely decaying turbulence \cite{KNNR-07, KKS}: quasi-singular distributions of di-vorticity are formed, which in $k$-space correspond to jets, leading to a strong turbulence anisotropy, then for times of the order of $10\tau$ turbulence becomes almost isotropic. In particular, at  these times any significant anisotropy in the angular fluctuations for the energy spectrum (for a fixed $k$) is not visible, while the probability distribution function of vorticity for large arguments has the exponential tail with the exponent linearly dependent on vorticity, in the agreement with the theoretical prediction \cite{FalkovichLebedev2011}.
\end{abstract}

\maketitle
\vspace{0.2 cm}
PACS: {52.30.Cv, 47.65.+a, 52.35.Ra}

\section{Introduction}

As was demonstrated in 1967 by Kraichnan \cite{kraichnan}, for the developed two-dimensional hydrodynamic turbulence in the inertial interval of scales there exist two Kolmogorov spectra, generated by two integrals of motion -- energy $E=1/2\int (\bf v)^2 d{\bf r}$ and enstrophy $1/2\int \Omega^{2}d\mathbf{r}$, where ${\bf v}$ - flow velocity and $\Omega =\mbox{rot}\,{\bf v}$ - vorticity. The first spectrum corresponds to a constant energy flux $\epsilon$ directed toward the region of small wave numbers (inverse cascade). This spectrum has the same dependence on $k$, as the famous Kolmogorov spectrum for three-dimensional hydrodynamic turbulence: $E(k)~\sim ~\epsilon^{2/3} k^{-5/3}$. The second spectrum - the Kraichnan spectrum
\begin{equation} 
E(k)~\sim ~\eta^{2/3}k^{-3} \label{Kraichnan}
\end{equation}
corresponds to a constant enstrophy flux $\eta$ towards the small-scale region (direct cascade). The existence of these two spectra has been confirmed in many numerical experiments simulating two-dimensional turbulence at high Reynolds numbers, i.e. in the regime, when in the leading  approximation, within the corresponding inertial intervals one can use the Euler equations,   instead of the Navier-Stokes equations  (see, e.g., \cite{boffetta} and  references therein). However, just after the Kraichnan paper \cite{kraichnan}, in the first numerical experiments \cite{lilly} there was observed  the emergence of sharp vorticity gradients corresponding to the formation of  jumps (quasi-shocks) with thickness  small compared to their length.  Based on these numerical observations, Saffman \cite{saffman} proposed another spectrum $E(k)~\sim ~k^{-4}$, the main contribution in which comes  from isotropically distributed quasi-shocks (in this meaning,  the Saffman spectrum is analogous to the Kadomtsev-Petviashvili spectrum \cite{KP} for acoustic turbulence). On the other hand, the Fourier amplitude from the vorticity jump  $\Omega _{k} \propto k^{-1}$, that immediately yields the Kraichnan type spectrum $E(k)~\sim ~k^{-3}$. However, the energy distribution from one such jump is anisotropic and has the form of a jet with an apex angle of the order of $\left( kL\right) ^{-1}$, where $L$ is characteristic length of the jump. It should be emphasized  that for isotropically distributed vorticity shocks we should arrive at the Saffman spectrum. In this sense  {\it the Kraichnan type spectrum generated by quasi-singularities must be anisotropic}. 
That was confirmed by both analytical arguments and numerical experiments in the case of a freely two-dimensional turbulence  when anisotropy in turbulence spectra is due to the presence of jets \cite{KNNR-07, KKS, K-04, KNNR-10}. In these papers, there was revealed the physical mechanism of quasi-shocks formation due to  a tendency to breaking (note that according to strong theorems \cite{wolibner} this process in a finite time is forbidden). This mechanism is associated with the property of frozenness  into fluid of the di-vorticity field  ${\bf B}=\mbox{rot}\,\Omega$ : in the inertial range of scales,  ${\bf B}$ obeys  the frozenness equation (see, e.g. \cite{Sulem, weiss}): 
\begin{equation} \label{B}
\frac{\partial \mathbf{B}}{\partial t} = \mbox {\rm rot}~[\mathbf{v}\times\mathbf{B}], \,\, \mbox{div}\,{\bf v}=0.
\end{equation}
From this equation, we see immediately that ${\bf B}$ can be changed  owing to the velocity component ${\bf v_n}$ normal to the force line of the field  ${\bf B}$. Because of the frozenness of the field ${\bf B}$  at the same time the component ${\bf v_n}$ represents the  velocity of the di-vorticity  line. In the general situation, for this component   $\mbox{div}\,
{\bf v}\neq 0$ that is a reason of compressibility for the field ${\bf B}$. 
The latter follows also from the fact that  equation (\ref{B}) permits integration in terms of the mapping $ {\bf r} = {\bf r} ({\bf a}, t) $, which is a  solution of the system for the Lagrangian trajectories defined by $ \mathbf{v}_{n}$:
$\cdot\mathbf{r}=\mathbf{v}_{n}(\mathbf{r},t);\quad
\mathbf{r}|_{t=0}=\mathbf{a}
$.
In this case, $\mathbf{B}$ is given by the following expression
\begin{equation} \nonumber
\mathbf{B}(\mathbf{r},t) = \frac{(\mathbf{B}_0(\mathbf{a})
\cdot\nabla_a)\mathbf{r}(\mathbf{a},t)}{J} ,
\end{equation}
where $\mathbf{B_0}(\mathbf{a})$  is the initial  field $\mathbf{B}$, and $J$ is the Jacobian of the mapping ${\bf r}={\bf r} ({\bf a}, t)$.  There are no any constrains imposed on $J$ so that it can take arbitrary values. In gas-dynamics, as known, a reason of breaking and formation of shock waves is connected with the compressibility of the transformation from the Eulerian description to the Lagrangian one with  vanishing of the corresponding mapping Jacobian. However, in the two-dimensional Euler hydrodynamics there is only  a tendency to form sharp vorticity gradients in the form of quasi-shocks, that was confirmed in numerical experiments on freely decaying turbulence \cite{KNNR-07,KKS,KNNR-10}. In particular, the increase of maximum value of $B$ in these experiments was 2 - 2.5 orders of magnitude, and the spatial distribution of $|B|$ was concentrated around  the lines (positions of quasi-shocks) with significantly less values of $|B|$ between these lines. 
In the energy spectrum, such quasi-shocks correspond to jets. Along of each jet the energy distribution decreases in accordance with the Kraichnan type power law $E\sim k^{-3}$.
Our  first results of numerical study of direct cascade for two-dimensional turbulence, i.e., in the presence  of both  pumping and dissipation, were presented in \cite{KuznetsovSereshchenko2015}. Pumping given by growth rate  $\Gamma(k)$ was concentrated in the small $k$ with a strong (singular at $k=0$) dissipation, providing suppression of inverse cascade. At large wave numbers, at $k>k_0 \sim 2/3 k_{max}$, we introduced a viscous type dissipation that allowed us to simultaneously solve the aliasing problem. 
At short times, in the inertial interval, turbulence  was developed by the same scenario as in the case of a freely decaying turbulence with the formation of both quasi-shocks and jets in the turbulence spectrum. 
In these experiments, at the initial stage we observed the formation of Kraichnan-type dependence of spectrum on the module $k$ ($E\sim k^{-3}$) at all angles, and the dependence of third-order velocity structure function $S_3=<\delta v_{\|}^3>$ on the separation length $R$ with strong anisotropy characteristic of the freely decaying turbulence. 
However, the  spectrum averaging over angles $E(k) = C_K\eta^{2/3}k^{-3}$, where $C_K\simeq 1.3$ is the Kraichnan constant,  coincided with the spectrum  previously obtained numerically in \cite{Gotoh}-\cite{LindborgVallgren}.  
It is important to note that the structure function $S_3$ averaging over angles gave the answer very different  from  its isotropic value. 
Analysis of these results testified in favor of the fact that the reason for this lies in the lack of spatial and temporal resolutions.
In this regard, we have been increased spatial resolution  up to $16384\times 16384$ and doubled the calculation time in comparison with the best experiments \cite{KuznetsovSereshchenko2015}. 
The main difference of the obtained results from the previous ones is that, at times of the order of $10\Gamma_{max}^{-1}$,  the jet structure of the spectrum in the direct cascade is destroyed  and turbulence tends  to be isotropic. In particular, on these times any significant anisotropy in angular fluctuations of energy spectrum (for a fixed value of $k$) is not observed. In the regime of an isotropic distribution, we found probability distribution functions $P$ for both  vorticity and  di-vorticity module $B$  The structure of the $P(\Omega)$ corresponds to the predictions of the isotropic theory  \cite{FalkovichLebedev2011}.

\section{Main equations and numerical scheme}

\begin{figure*}[]
	\centerline{ 
		\includegraphics[width=0.3\textwidth]{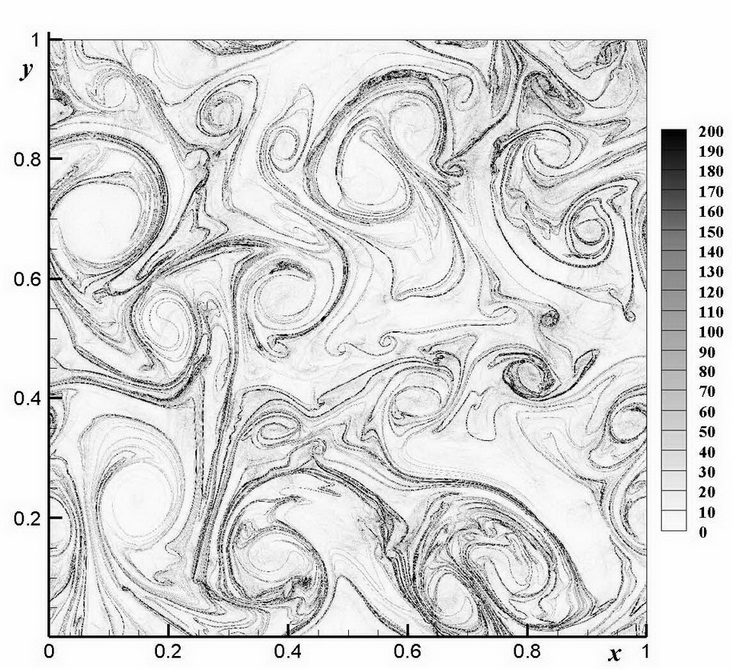}
		\includegraphics[width=0.3\textwidth]{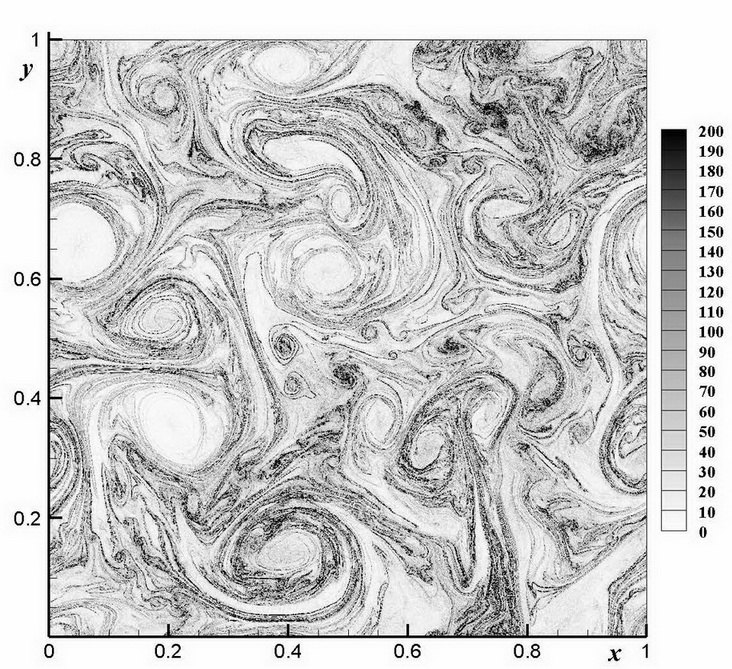}
		\includegraphics[width=0.3\textwidth]{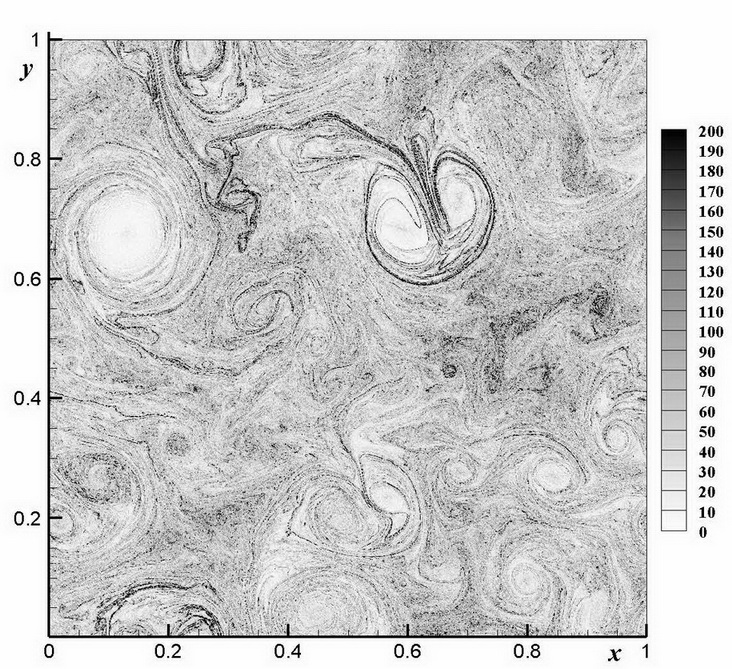}}
	\begin{tabular}{p{0.42\linewidth}p{0.13\linewidth}p{0.42\linewidth}}
		\centering a) & \centering b) & \centering c) \\
	\end{tabular}
	\caption{Fig.1. Distribution of $|B|$ at $t=150,250,450$.}
	\label{fig1}
\end{figure*}
Let us briefly consider the equations of motion and the numerical scheme, which completely coincide with those in \cite{KuznetsovSereshchenko2015}. 

Numerical simulation of the direct cascade for two-dimensional turbulence  in the framework of the equation for the vorticity $\Omega =\nabla \times \mathbf{u}$ was performed in a square box with the size  $L=1$
with periodic boundary conditions along both coordinates, where $\mathbf{u(x},t)$ is a fluid velocity satisfying  the incompressibility condition  $\emph{div}\,\mathbf{u}=0$:
\begin{equation}
\frac{\partial \Omega }{\partial t}+(\mathbf{u}\nabla )\Omega =\hat{\Gamma}\Omega +\hat{\gamma}\Omega.
 \label{NS}
\end{equation}%
Here the operator $\hat{\Gamma}$ is responsible for both injection of the energy and its dissipation on large scales, to prevent reverse cascade, and the operator $\hat{\gamma}$ for dissipation enstrophy at large $k$. Both of these operators were set by their Fourier transforms (see \cite{KuznetsovSereshchenko2015}):
\begin{eqnarray}
\Gamma _{k} &=&A\frac{(b^{2}-k^{2})(k^{2}-a^{2})}{k^{2}}\quad \mbox{at}\quad
0\leq k\leq b,\quad \nonumber \\
\quad \Gamma _{k} &=&0\quad \mbox{at}\quad k>b, \nonumber
\end{eqnarray}
and 
\begin{eqnarray}
\gamma _{k} &=&0\quad \mbox{\rm at}\quad k\leq k_{c},\quad \nonumber \\
\quad \gamma _{k} &=&-\nu (k-k_{c})^{2}\quad \mbox{\rm at}\quad k>k_{c}. \nonumber
\end{eqnarray}
In the numerical integration of the equation (\ref{NS}) parameters $a$ and $b$ are chosen from the conditions of the most rapid transition of the system to the steady-state regime at small $k$. Below are the results with $A=0.004$, $a=3$ è $b=6$. For dissipation in the viscous-type form, providing enstrophy absorption, coefficient of viscosity was $\nu=1.5$ and $k_{c}$  was $0.6k_{\max }$, where $k_{\max}=8192$, that simultaneously solves the problem of aliasing. The initial conditions were the same as in our previous papers \cite{KKS,KuznetsovSereshchenko2015}, as random set of vortices of the Gaussian shape, randomly distributed over the entire domain with zero mean vorticity. Equation (\ref{NS}) was solved numerically  as well as in our previous works \cite{KKS,KuznetsovSereshchenko2015} by means of   pseudospectral Fourier method, while integration in time was performed with the use of a hybrid Runge--Kutta/Crank–Nicholson third-order scheme. Numerical experiments \cite{KuznetsovSereshchenko2015} performed on the grid with a maximum resolution of $8192\times 8192$, in this paper we present the results of numerical experiments on the grid $16384\times 16384$. Simulations were performed (with the use of the NVIDIA CUDA technology) at the Computer Center of the Novosibirsk State University.

\section{Numerical results}

At the initial stage, for the times of order of the inverse  pumping growth rate $\Gamma_{max}^{-1}$, the development of turbulence is about the same scenario as in the case of a freely decaying turbulence \cite{KKS}:  quasi-singular distributions of di-vorticity are formed, which in $k$-space correspond to jets, leading to a strong turbulence anisotropy. Fig. 1a shows a typical distribution of the di-vorticity module $|B|$, which is the most concentrated on the lines (positions of quasi-shocks). Between these lines the value of $|B|$ is significantly lower. Accordingly, jets (with weak and strong overlapping in the k-space) are observed in the spectrum, as a result, the turbulence spectrum has a large anisotropic component. Fig. 2à shows in k-space the distributions of the energy density of fluctuations  $\epsilon({\bf k})$, normalized by $k^{-4}$
\begin{figure*}[t]
\centerline{
\includegraphics[width=0.3\textwidth]{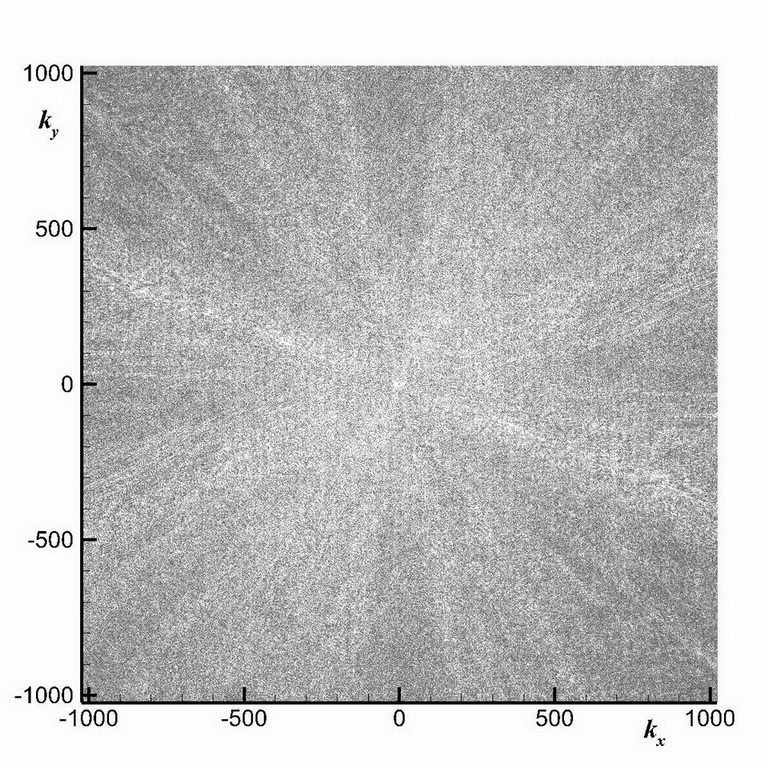}
\includegraphics[width=0.3\textwidth]{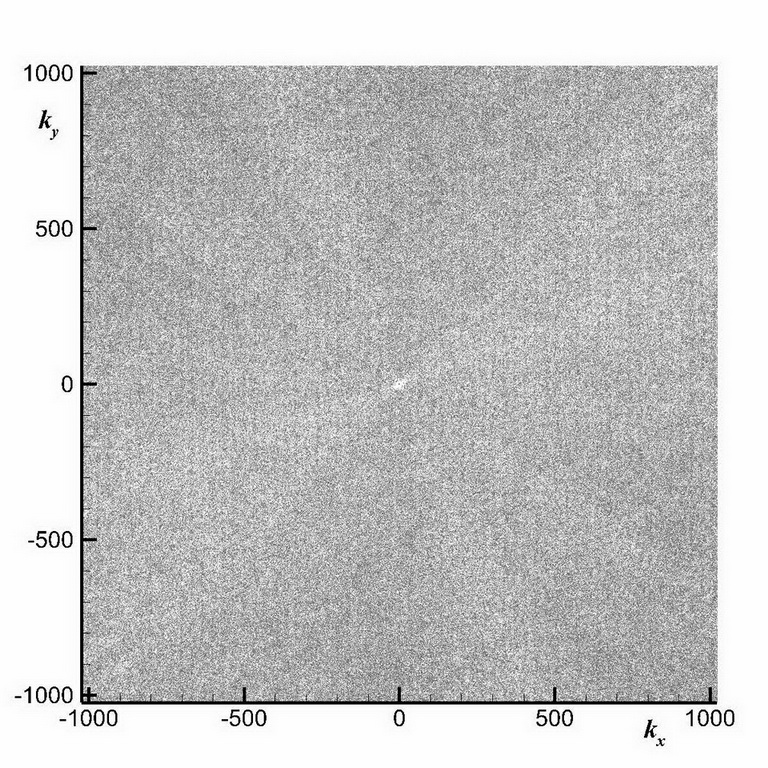}
\includegraphics[width=0.3\textwidth]{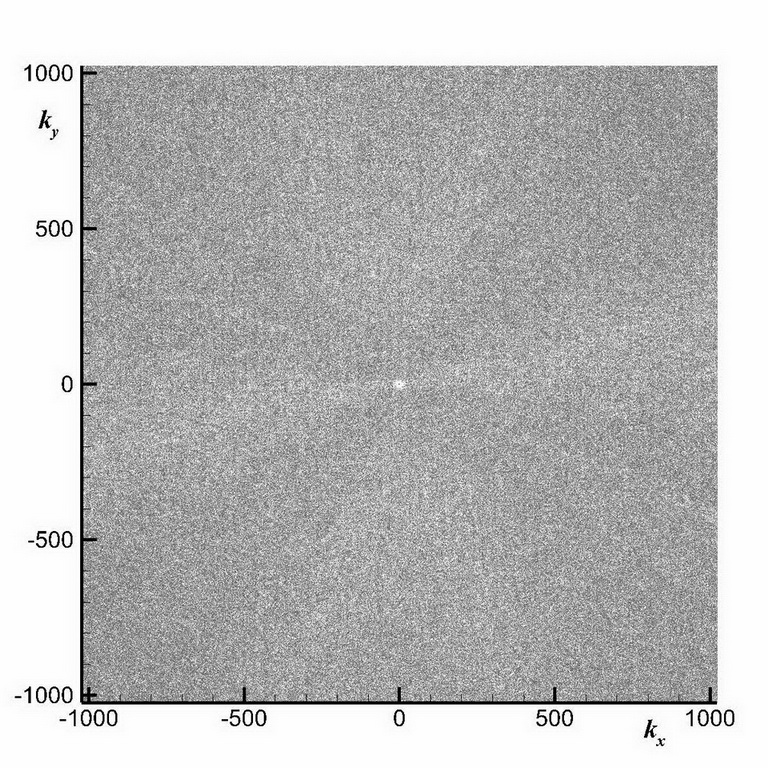}}
\begin{tabular}{p{0.42\linewidth}p{0.13\linewidth}p{0.42\linewidth}}
\centering à) & \centering á) & \centering â) \\
	\end{tabular}
	\caption{Fig.2.  Distributions  of the energy density of fluctuations $\epsilon({\bf k})$ (normalized by $k^{-4}$) at $t=150, 250, 450$.}
	\label{fig2}
\end{figure*}
At each angle in the $k$-space  in the inertial interval value of $\epsilon({\bf k})k^{4}$ at a given time fluctuates greatly, and after averaging is almost constant. So, after multiplying the $\bar{\epsilon({\bf k})} $ to $k$ we obtain the angular dependence of energy spectrum $ E(k,\theta)\sim k^{-3}$ (Fig. 3a shows a typical dependence of $\epsilon({\bf k})k^{4}$ for angle $\phi=45^{\circ}$ at $t=150$, Fig. 3b - the result of the averaging over the fluctuations of this value). 
It is important to note that the formation of the Kraichnan-type dependence on the module $k$ occurs on the first stage of the direct cascade development, when enstrophy transfer reaches the "viscous" region. According to our estimates, the time of such transfer is order of the inverse pumping growth rate $\Gamma_{max}^{-1}$. At this stage, the energy spectrum depends strongly on the angle. It is surprising that after averaging over the angles spectrum $E(k)$ having both the Kraichnan-type dependence on $k$ and enstrophy flux $\eta$, defined as $1/2\int \gamma(k) |\Omega_k |^2 d{\bf k}$, gives a value for the Kraichnan constant $C_K\simeq 1.3$ which coincides with that previously obtained in numerical experiments \cite{Gotoh}-\cite{LindborgVallgren}.
\begin{figure}[t]
\label{fig3}
\centerline{
\includegraphics[width=0.23\textwidth]{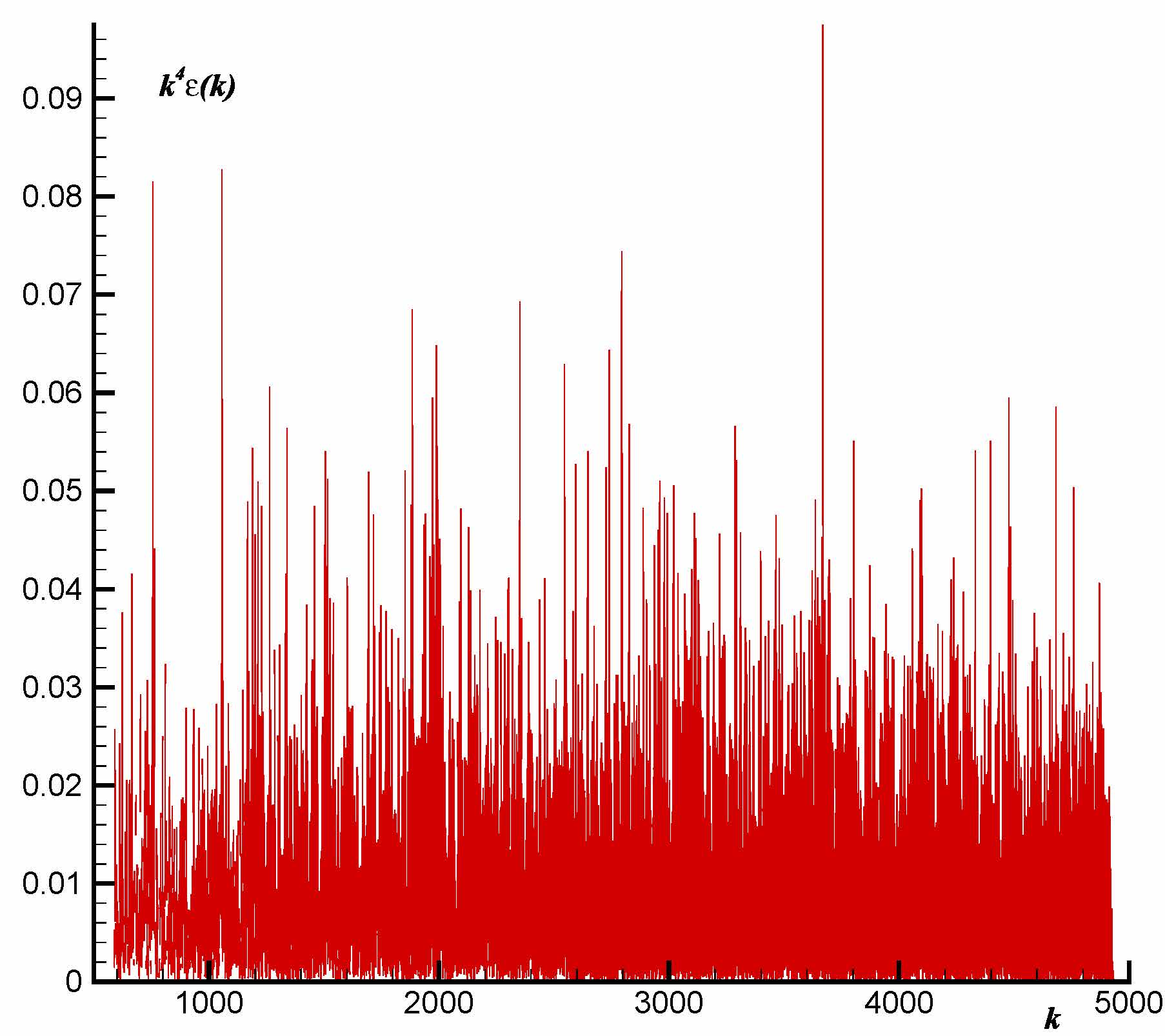}
\includegraphics[width=0.23\textwidth]{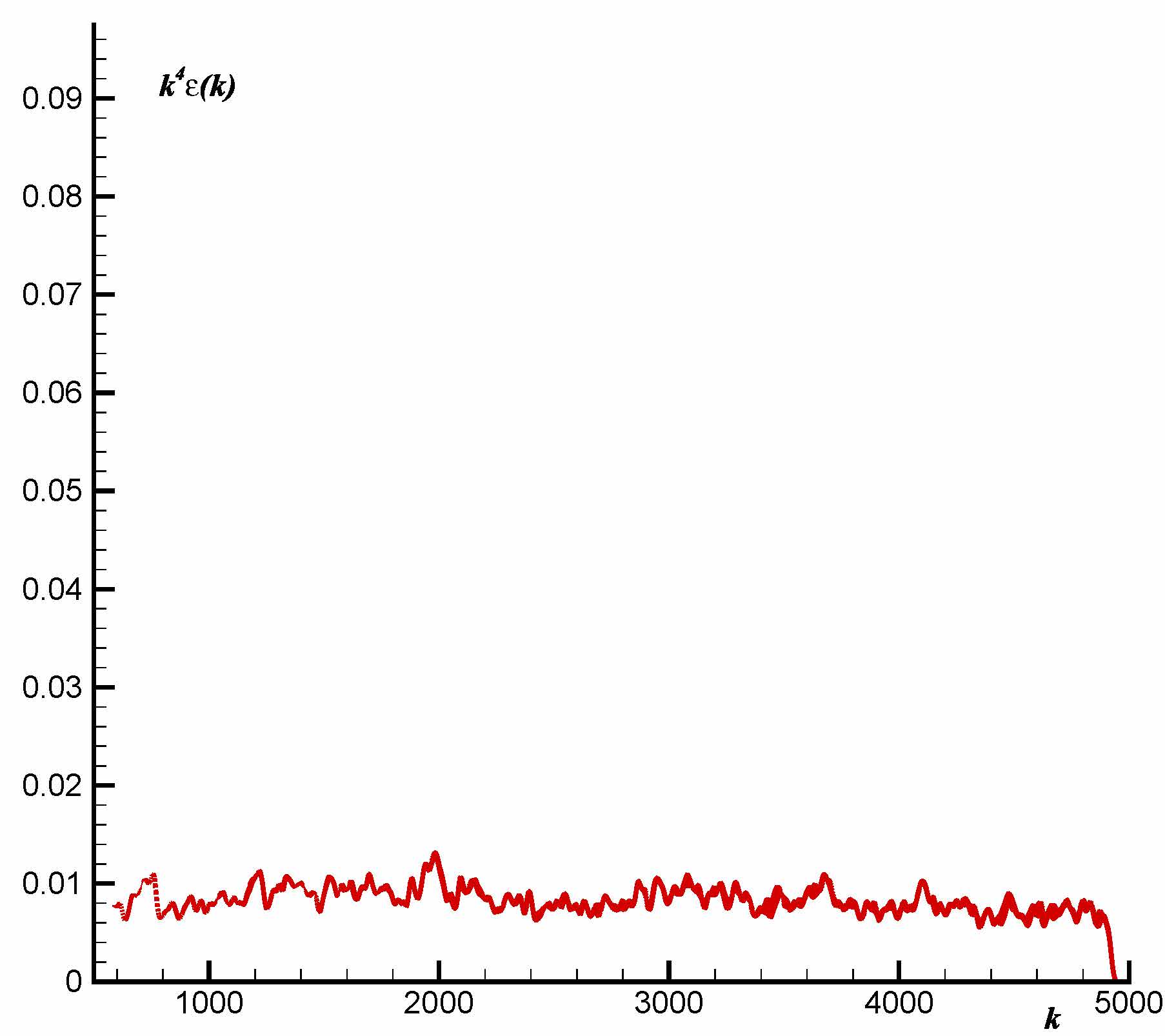}}
\begin{tabular}{p{0.50\linewidth}p{0.40\linewidth}}
\centering a) & \centering b)\\
\end{tabular}
\caption{Fig.3. Dependencies of $\epsilon({\bf k})k^{4}$ before averaging (a) and averaging (over domain $\Delta k=100$) value of $\bar{\epsilon({\bf k})k^{4}}$  (b) on $k$ for angle $\phi=45^{\circ}$ at $t=150$.}
\end{figure}
\begin{figure}[t]
\label{fig4}
\centerline{
\includegraphics[width=0.45\textwidth]{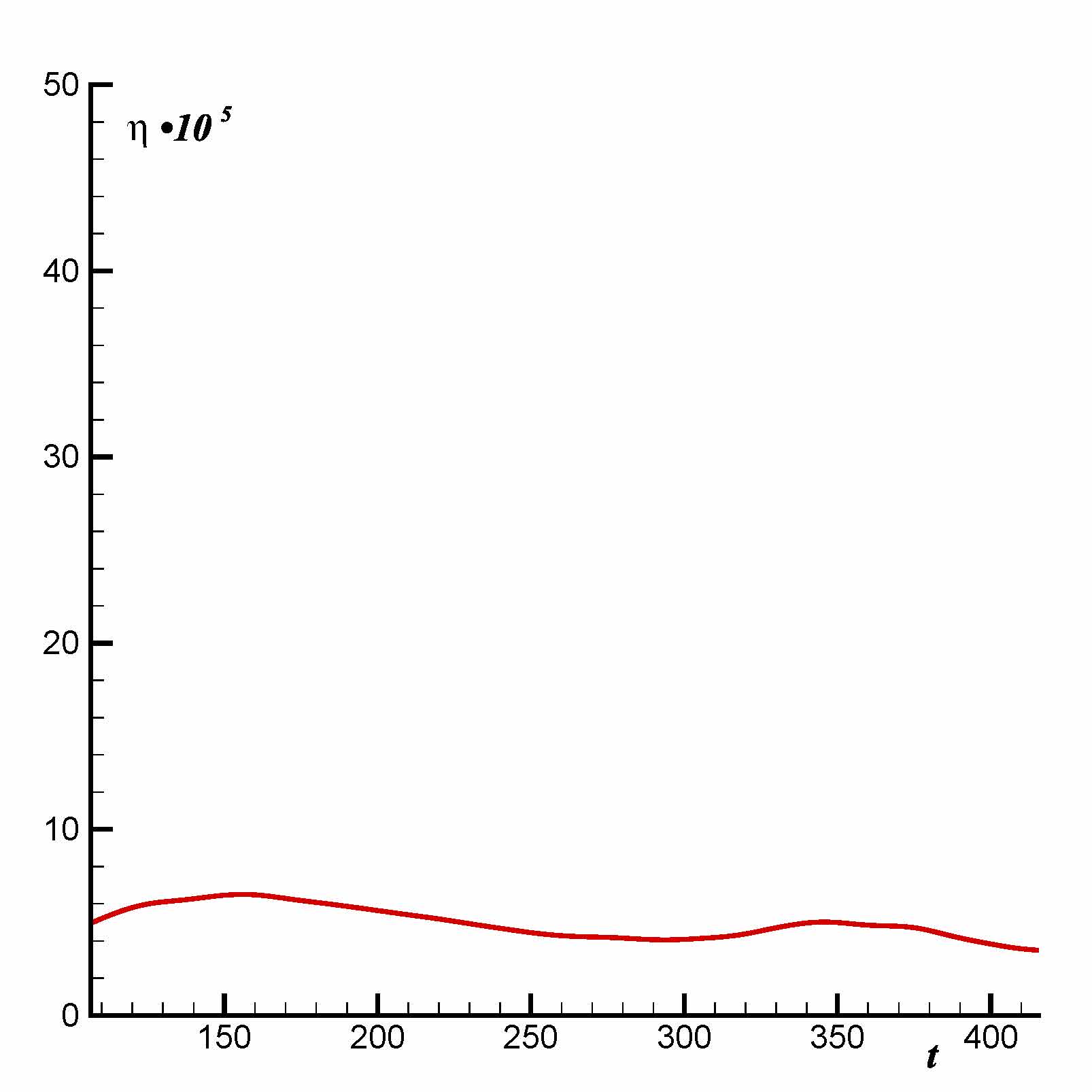}}
\caption{Fig.4. Temporal dependence of the enstrophy flux $\eta$.}
\end{figure}
In the next stage, the net of quasi-shocks lines becomes  more complicated (turbulent) (Fig.1b). The distances between quasi-shocks lines are reduced, and as a result the anisotropy in energy spectrum decreases (Fig. 2b). Finally, for times of the order of $10 \Gamma_{max}^{-1}$ jets practically disappear (Fig. 1c and Fig. 2c) and turbulence in the direct cascade becomes almost isotropic. It also appears that at all times, starting from occurrence of jets up to their disappearance, enstrophy flux is almost constant (Fig. 4). The total energy of sufficiently fast becoming a constant (at the first stage), which is not for the total enstrophy. It is close to a constant value only at the isotropization stage (Fig. 5).
\begin{figure}[t]
\label{fig5}
\centerline{
\includegraphics[width=0.23\textwidth]{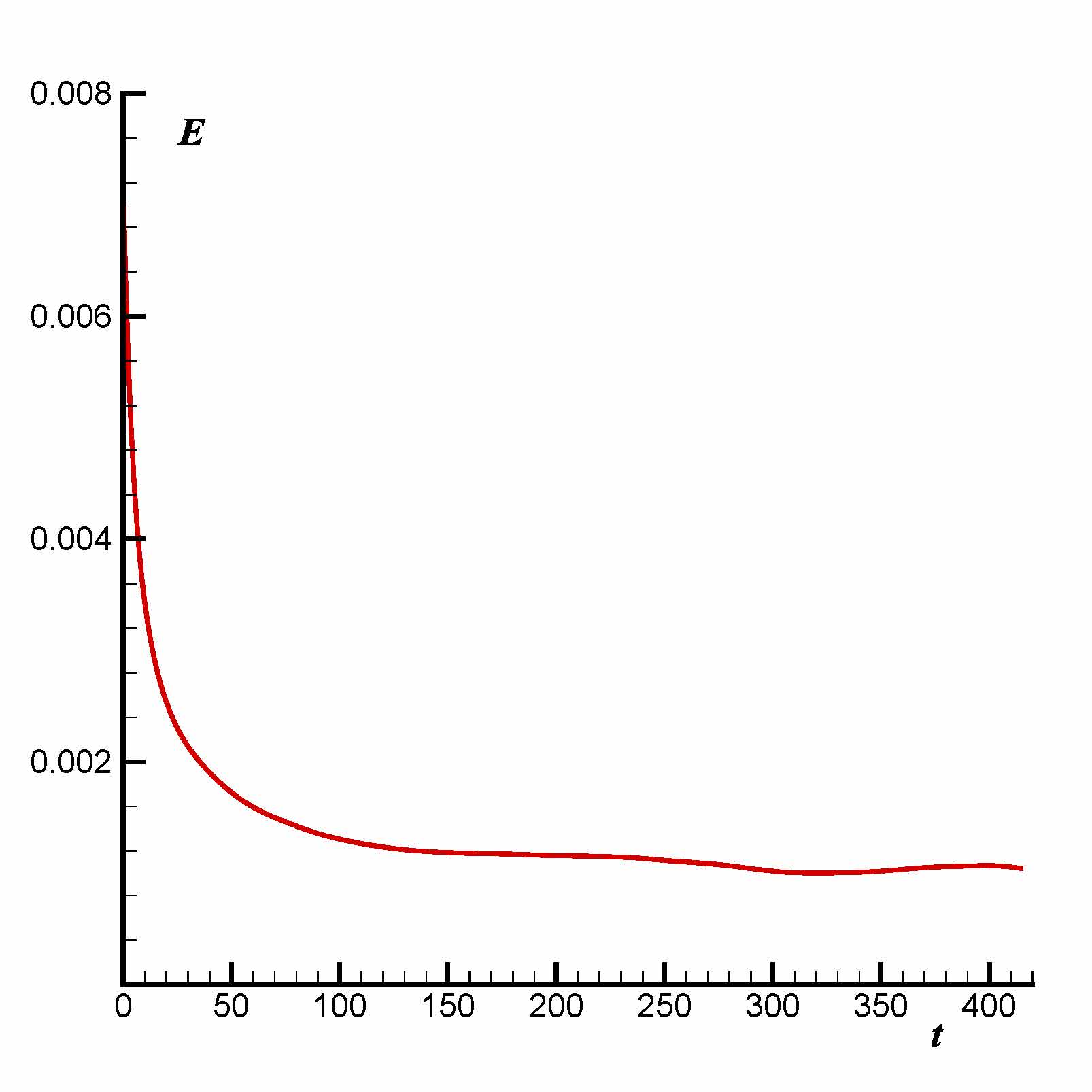}
\includegraphics[width=0.23\textwidth]{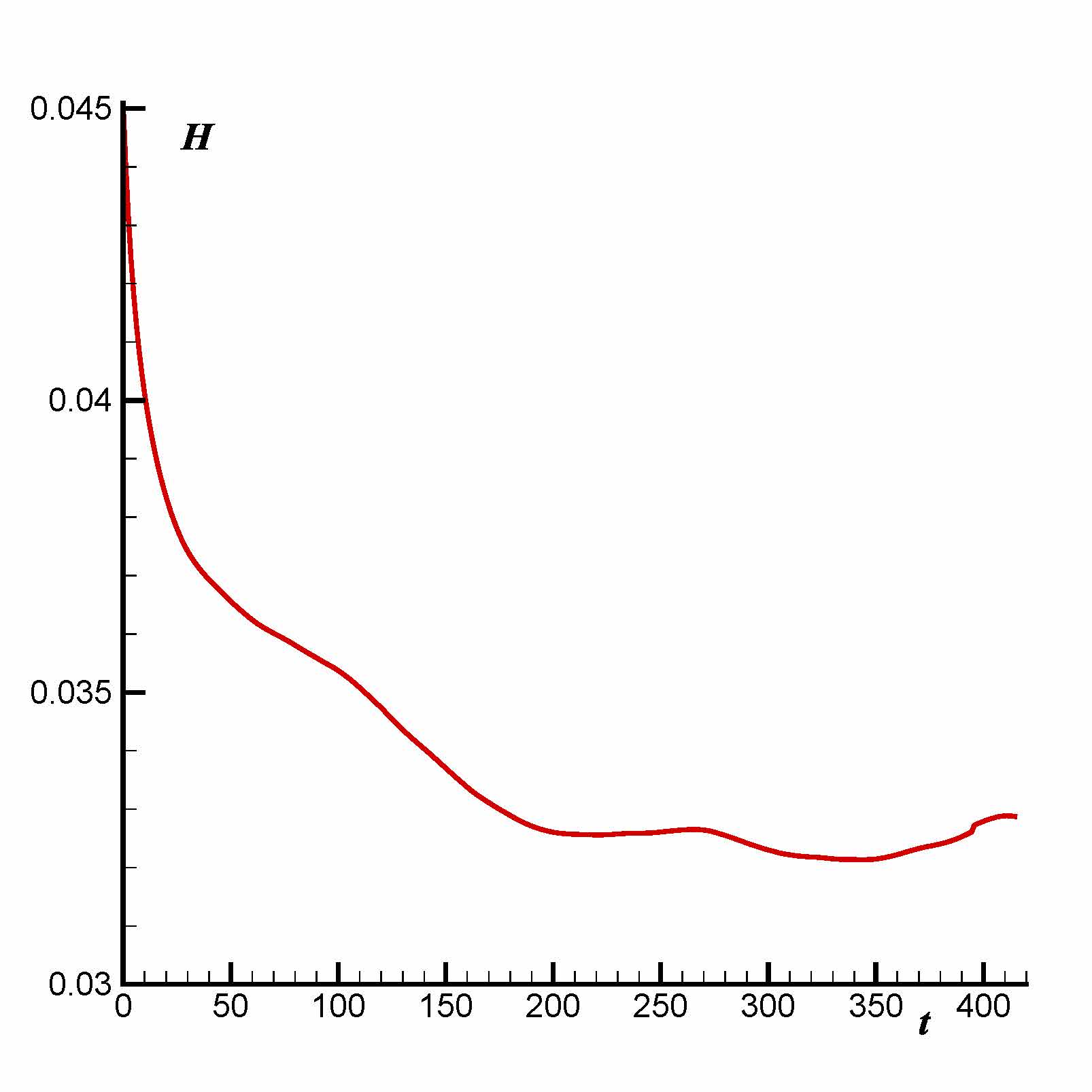}}
\caption{Fig.5. Time dependences of the total energy $E$ and the total enstrophy $H$.}
\end{figure}
Another indication of turbulence isotropization at times of the order of $10 \Gamma_{max}^{-1}$  found in this paper is a isprobability distribution function of vorticity $P$ (Fig. 6), which for large arguments has an exponential tail with exponent $\beta$, linearly dependent on vorticity $\bar {\Omega}$, in agreement with the theoretical predictions \cite{FalkovichLebedev2011}. According to these predictions, the angle slope of the exponent  is order of ${\bar\Omega}_{rms}^{-1}$, where $\bar{\Omega}_{rms}=[\eta* ln(L/R)]^{1/3}$ is rms vorticity fluctuations. Numerical experiment (Fig. 6) gives the asymptotic behavior of $P=0.0005exp(-2.3{\bar\Omega})$ with ${\bar\Omega}_{rms}=0.43$.  If calculate the exstrophy flux as the integral $\eta=1/2\int \gamma|\Omega_k|^2d{\bf k}$, then $\bar{\Omega}_{rms}=0.15$. Calculation of $\bar{\Omega}_{rms}$ by a given distribution function gives the value of $0.2566$. Thus, the values $\bar{\Omega}_{rms}$ are close to each other with accuracy of the order of unity.
\begin{figure}[t]
\label{fig6}
\centerline{
\includegraphics[width=0.45\textwidth]{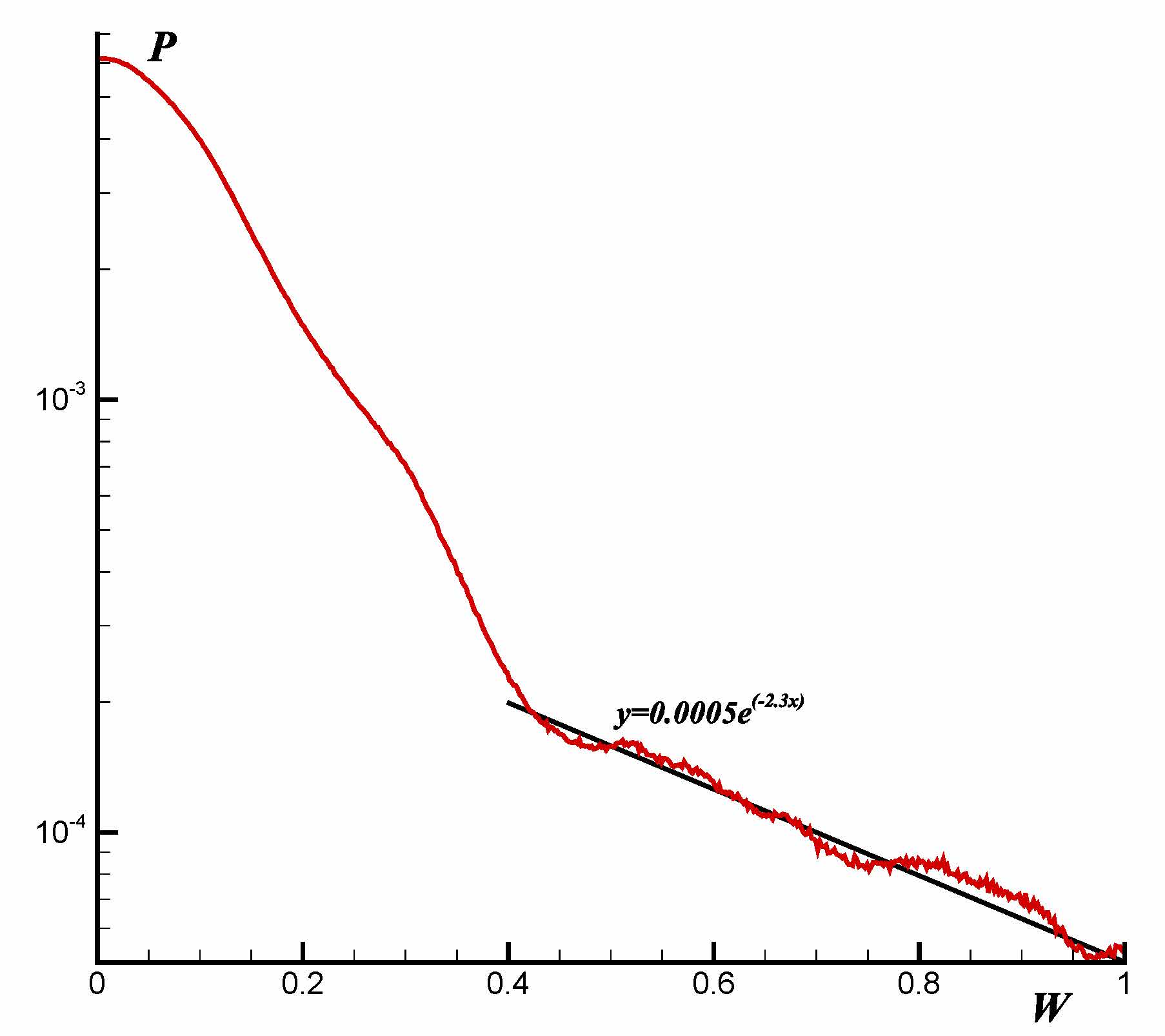}}
\caption{Fig.6. Probability distribution function $P$ for vorticity  at $t=450$.}
\end{figure}
The corresponding distribution function $P$ for di-vorticity $B$ also has two specific regions (Fig. 7): in the first one the distribution function is close to the Poisson distribution, $\sim B\exp (-B^2/B_0^2)$, in the second region (large value of di-vorticity $B$) the distribution function $P$ is of exponential behavior with more pronounce  linear dependence of exponent on $B$ than the similar one for vorticity. For this numerical experiment  $B_{rms}$, calculated from the angle slope, is equal to $88$. If calculate the $B_{rms}$ by means of the distribution function $P(B)$, then this value is equal to  $84.6$. 
\begin{figure}[t]
\label{fig7}
\centerline{
\includegraphics[width=0.45\textwidth]{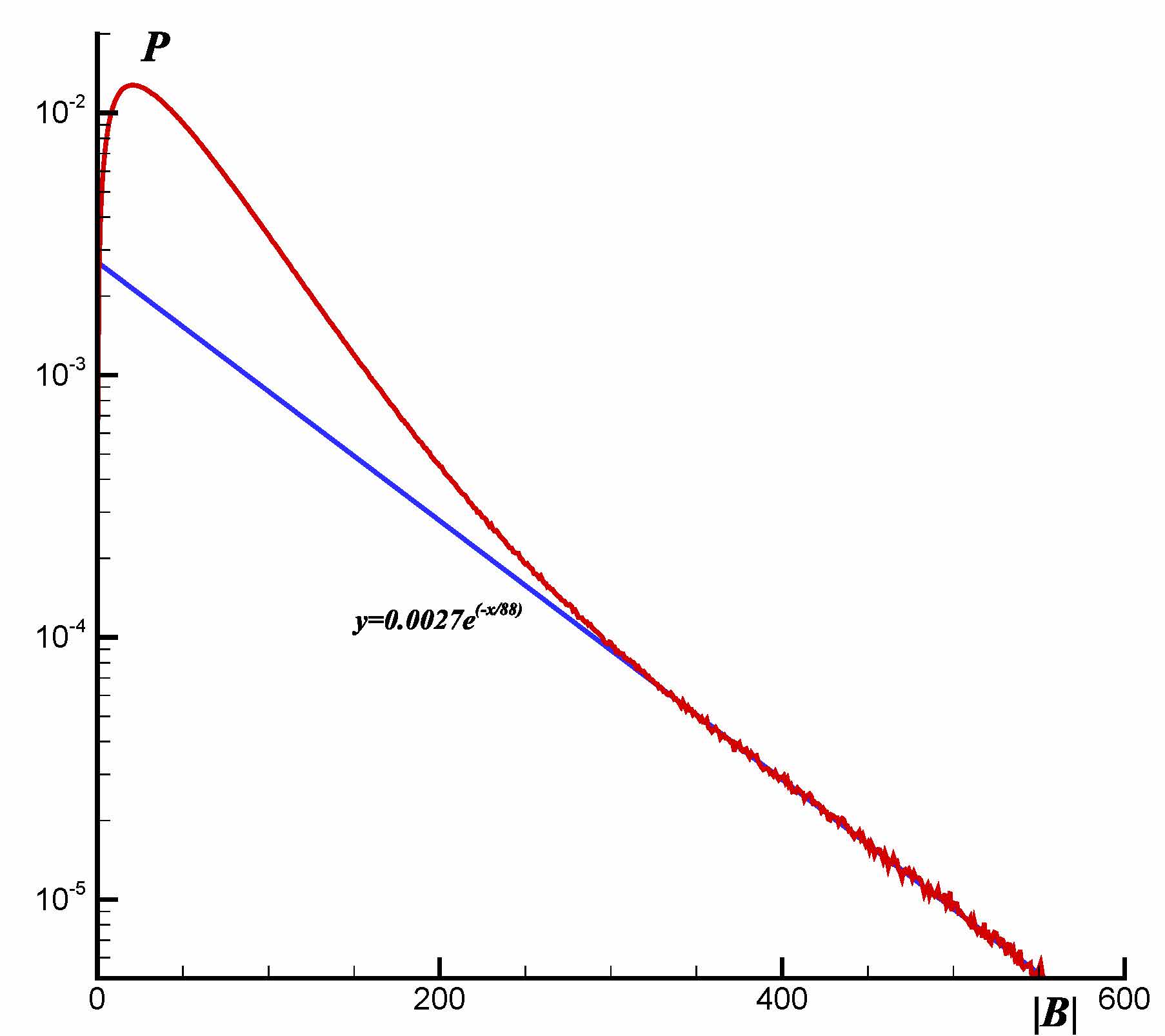}}
\caption{Fig.7. Probability distribution function of di-vorticity $P$ at $t=450$.}
\end{figure}

\section{Conclusion}

The main conclusion of this paper is that in the direct cascade the formation of a power-law dependence on wave number $k$ for the turbulence spectrum  with the Kraichnan exponent due to the vorticity quasi-shocks  represents the fastest process. 
In this stage, in the turbulence spectrum there is a strong anisotropy owing to jets which are the Fourier transforms of quasi-shocks. In the next -- much slower -- stage, the structure of quasi-shocks lines becomes more complicated (turbulent). The distances between quasi-shocks lines are reduced, and the spectrum tends to  more isotropic. It should be  also noted that at these times for the distribution probability function of vorticity we have observed the formation of the exponential tail at large arguments with the exponent which can be extrapolated as a linear function of vorticity in agreement with the theoretical predictions   \cite{FalkovichLebedev2011}.
  The probability distribution function $P$ for di-vorticity $B$ also has two specific regions: in the first region the distribution function is close to the Poisson distribution, $\sim B\exp (-B^2/B_0^2)$, in the second one (large value of di-vorticity $B$) distribution function $P$ is of exponential behavior with more pronounce linear dependence on $B$ than the similar one for vorticity. Both of these observations suggest that direct cascade of turbulence at large times loses anisotropy due to a tendency to breaking. In our opinion, there are at least two possible reasons of the turbulence isotropization.

The first reason may be related to the pumping area, where, in spite of the strong dissipation at low $k$, large-scale vortices are formed (these are some remains not killed until the end of inverse cascade), which, due to their rotation, contribute into the system of vorticity quasi-shocks additional stretching of di-vorticity lines, and, on the other hand,it makes the system of significant lines of the field $B$ more complicated.

Another possible reason is connected with interaction of turbulence with the viscous type region. We have  observed that the spectrum isotropization happens for characteristic  times much larger the Kraichnan enstrophy transferring time, when the enstrophy flux wave reaches the viscous region. As is known, the direct cascade of turbulence is a non-local (or rather - weakly nonlocal, that is accompanied by the appearance of logarithmic corrections to the Kraichnan spectrum $E(k)\sim k^{-3}$, see \cite{kraichnan} and \cite{FalkovichLebedev1994}). Locality of turbulence means that the main nonlinear interaction is interaction between scales of the same order. Interaction between very different scales is strongly depressed. In this situation, both boundaries to the inertial interval, i.e. pumping and viscous dissipation areas, as isotropic sources, in our opinion, are responsible for the turbulence isotropization of direct cascade.

\vspace{0.5cm}
We are grateful to V.V. Lebedev and I.V. Kolokolov for useful discussions.
This work was supported by the Russian Science Foundation (project no. 14-22-00174).

\end{document}